# Giant orbital Hall effect and orbital-to-spin conversion in 3*d*, 5*d*, and 4*f* metallic heterostructures

Giacomo Sala [*] and Pietro Gambardella [†]
*Department of Materials, ETH Zurich, 8093 Zurich, Switzerland*

The orbital Hall effect provides an alternative means to the spin Hall effect to convert a charge current into a flow of angular momentum. Recently, compelling signatures of orbital Hall effects have been identified in 3*d* transition metals. Here, we report a systematic study of the generation, transmission, and conversion of orbital currents in heterostructures comprising 3*d*, 5*d*, and 4*f* metals. We show that the orbital Hall conductivity of Cr reaches giant values of the order of $5 \times 10^5$ $[\frac{\hbar}{2e}]$ $\Omega^{-1}$ m$^{-1}$ and that Pt presents a strong orbital Hall effect in addition to the spin Hall effect. Measurements performed as a function of thickness of nonmagnetic Cr, Mn, and Pt layers and ferromagnetic Co and Ni layers reveal how the orbital and spin currents compete or assist each other in determining the spin-orbit torques acting on the magnetic layer. We further show how this interplay can be drastically modulated by introducing 4*f* spacers between the nonmagnetic and magnetic layers. Gd and Tb act as very efficient orbital-to-spin current converters, boosting the spin-orbit torques generated by Cr by a factor of 4 and reversing the sign of the torques generated by Pt. To interpret our results, we present a generalized drift-diffusion model that includes both spin and orbital Hall effects and describes their interconversion mediated by spin-orbit coupling.

## I. INTRODUCTION

The interconversion of charge and spin currents underpins a variety of phenomena and applications in spintronics, including spin-orbit torques, spin pumping, the excitation of magnons, and the tuning of magnetic damping [1,2]. The spin Hall effect (SHE) mediates this interconversion through the combination of intrinsic and extrinsic scattering processes, all of which require sizable spin-orbit coupling [3]. Recent theoretical work has shown that the intrinsic SHE is accompanied by a complementary process involving the orbital angular momentum, the so-called orbital Hall effect (OHE), which consists in the flow of orbital momentum perpendicular to the charge current [4–10]. According to theoretical calculations, the OHE is more common and fundamental than the SHE because it does not require spin-orbit coupling and can thus occur in a wider range of materials. The intrinsic SHE then emerges as a by-product of the OHE resulting from the orbital-to-spin conversion in materials with nonzero spin-orbit coupling. In this case, the spin Hall conductivity has the same sign as the product between the orbital conductivity and the expectation value of spin-orbit coupling: $\sigma_S \sim \sigma_L \langle \mathbf{L} \cdot \mathbf{S} \rangle$. The OHE was first predicted in 4*d* and 5*d* transition elements [11,12] and recently in light metals [4] and their interfaces [13] as well as in two-dimensional (2D) materials [14,15]. The theoretical orbital Hall conductivity of light elements is comparable to or even larger than the spin Hall conductivity of Ta, W, and Pt, which provide a strong SHE [4]. The OHE is thus intrinsically more efficient than the SHE, and orbital currents are expected to contribute to magnetotransport effects such as the anisotropic, spin Hall, and unidirectional magnetoresistance as well as spin-orbit torques [6,16–20]. The ubiquity and strength of the OHE, besides making it fundamentally interesting, broaden the material palette available for spintronic applications and provide an additional handle to optimize the efficiency of spin-orbit torques. Yet, differently from spins, nonequilibrium orbital currents do not couple directly to the magnetization of magnetic materials and can torque magnetic moments only indirectly through spin-orbit coupling [19–21]. Optimizing the orbital-to-spin conversion is thus a prerequisite for taking advantage of large orbital currents.

The prediction of the OHE in light elements has triggered intense research on current-induced orbital effects. Recent experiments have identified signatures of the OHE in materials with low [18,19] and high [20] spin-orbit coupling and revealed its contribution to spin-orbit torques [16,18], whose strength can be tuned by improving the orbital-to-spin conversion ratio [19,22]. However, experimental values of the orbital Hall conductivity are smaller than theoretical estimates [16,18,19,23], and a systematic investigation of orbital effects as a function of the type and thickness of nonmagnetic, ferromagnetic, and spacer layers is still missing.

Here, we present a comprehensive study of the interplay of the OHE and SHE in structures combining different light and heavy nonmagnetic metals (NM = Cr, Mn, Pt), ferromagnets (FM = Co, Ni), and rare-earth spacers (X = Gd, Tb). We provide evidence of the OHE in Pt and Mn and report giant values of the orbital Hall conductivity in Cr, which extrapolate to the theoretical limit of $10^6 \, [\frac{\hbar}{2e}] \, (\Omega \, \text{m})^{-1}$ in Cr films thicker than the orbital diffusion length [4], which we estimate to be $\gtrsim 20$ nm. Because of the simultaneous presence of strong OHE and SHE in Pt and Cr, we argue that experimental results are best described by a combined spin-orbital conductivity rather than by separating the two effects. We show that the interplay between orbital and spin currents can be tailored by varying the thickness of the ferromagnetic layer as well as by inserting a Gd or Tb conversion layer between the nonmagnet and the ferromagnet. Rare-earth spacers do not generate significant spin-orbit torques by themselves, but they enhance the torque efficiency up to four times when Cr is the source of spin and orbital currents and reverse the sign of the torques generated by Pt. The latter effect is attributed to the OHE overcoming the SHE in Pt. Finally, we present a phenomenological extension of the spin drift-diffusion model that includes orbital effects and the conversion between spin and orbital moments, which accounts for both the thickness dependence and sign change of the spin-orbit torques generated by the interplay of OHE and SHE in NM/FM and NM/$X$/FM heterostructures.

[*]giacomo.sala@mat.ethz.ch
[†]pietro.gambardella@mat.ethz.ch

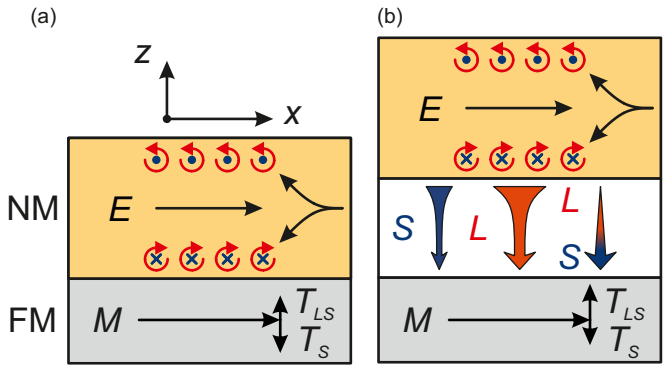

FIG. 1. (a) The spin Hall effect and orbital Hall effect induced by an electric field $E$ in a nonmagnet (NM) produce spin ($T_S$) and orbital ($T_{LS}$) torques on the magnetization $M$ of an adjacent ferromagnet (FM). The strength of the torques depends on the intensity of the spin and orbital currents and on the spin-orbit coupling of the ferromagnet. The schematic shows the direction of the induced spin ($S$, blue dots and crosses) and orbital ($L$, red circling arrows) *angular momenta* when the spin and orbital Hall conductivities $\sigma_{S,L} > 0$. (b) The insertion of a spacer layer may increase the orbital torque relative to the spin torque by converting the orbital current (red) into a spin current (blue) prior to their injection into the ferromagnet.

## II. BACKGROUND

According to the theory of the OHE, an electric field applied along the $x$ direction in a material with orbital texture in $k$ space induces interband mixing that results in electron states with finite orbital angular momentum [5–7]. Electrons occupying these nonequilibrium states carry the angular momentum as they travel in real space. Therefore, although the total orbital momentum vanishes, a nonzero orbital current is produced along the $z$ ($y$) direction with orbital polarization parallel to $\pm y$ ($\pm z$), similar to the SHE [see Fig. 1(a)]. The latter occurs concomitantly with the OHE when the nonmagnet has nonzero spin-orbit coupling $\langle \mathbf{L} \cdot \mathbf{S} \rangle_{\text{NM}}$. The primary spin current injected into the adjacent ferromagnet exerts a direct torque on the local magnetization (spin torque). Orbitals, instead, act indirectly through the spin-orbit coupling of the ferromagnet that converts the orbital current into a secondary spin current. We refer to the torque generated by this secondary spin current as orbital torque. The (independence) dependence of the (spin) orbital torque on $\langle \mathbf{L} \cdot \mathbf{S} \rangle_{\text{FM}}$ is the key difference between SHE and OHE. In the SHE scenario, the angular momentum is entirely generated in the nonmagnet, and the ferromagnet behaves almost as a passive layer since it only contributes to the properties of the NM/FM interface. In contrast, the OHE in a NM/FM bilayer depends on both the interfacial and bulk properties of the ferromagnet, which is directly involved in the torque generation. Since the orbital conductivity is typically large [$\approx 10^5 \, (\Omega \, \text{m})^{-1}$] [4] but the spin-orbit coupling of 3$d$ ferromagnets is relatively weak [24], the orbital torque efficiency in NM/FM bilayers is finite but small. Alternatively, the orbital torque may be enhanced by realizing most of the orbital-to-spin conversion in a spacer layer sandwiched between the nonmagnet and the ferromagnet [Fig. 1(b)]. The effectiveness of this approach depends on the conversion efficiency of the spacer, its spin and orbital diffusion lengths, and the quality of the additional interfaces, as discussed later.

Here, we summarize fundamental theoretical predictions and experimental confirmations of the OHE. We list approaches to distinguish orbital and spin effects by means of torque measurements in heterostructures with different elements, thickness, and stacking order. Furthermore, we establish a parallel between known spin-transport effects and possible orbital counterparts that have not been observed yet but could contribute to answering open questions about orbital transport.

(i) Large orbital Hall conductivities have been predicted in several 3$d$, 4$d$, and 5$d$ transition elements [4,11,12] and 2D materials [9,10]. Experimental evidence is so far limited to Cr [19,25], Cu [20,21,26], Zr [18], and Ta [20]. Recent experiments on V [23,27] can also be reinterpreted in light of the OHE. The coexistence of the OHE and the SHE, especially in heavy metals, makes it difficult to distinguish the two effects.

(ii) The spin and orbital torques are expected to add constructively (destructively) when $\langle \mathbf{L} \cdot \mathbf{S} \rangle_{\text{NM}} \cdot \langle \mathbf{L} \cdot \mathbf{S} \rangle_{\text{FM}} > 0$ ($<0$). This competition can be tailored by properly choosing the ferromagnet, as recently observed in Refs. [19,20].

(iii) In a NM/FM bilayer, the orbital Hall efficiency should depend on the thickness of both the nonmagnet ($t_{\text{NM}}$) and the ferromagnet ($t_{\text{FM}}$). In contrast, the spin Hall efficiency is nominally independent of the latter and results in an inverse dependence of the spin torque on $t_{\text{FM}}$ [1]. The dependence of the orbital Hall efficiency on $t_{\text{NM}}$ has been addressed in Ref. [19], but the role of $t_{\text{FM}}$ is still unknown.

(iv) The spin diffusion in transition metals with strong SHE is typically limited to a few nanometers [28]. Although recent measurements suggest longer orbital diffusion lengths [19,29], the length scale of the orbital diffusion and its conversion into spins remain to be established. These quantities and the nature of the mechanisms underlying the orbital scattering may be ad-dressed by torque measurements in thick nonmagnetic films and by nonlocal transport measurements, which could also verify the existence of the inverse OHE.

(v) Spacer layers between the nonmagnet and the ferromagnet can alter spin torques in several ways, namely, by introducing an additional interface with different spin scattering properties, by suppressing the spin backflow, and by modifying the spin memory loss [1,30–33]. Such effects are expected to influence also the orbital torque. In addition, spacers can either increase or decrease the orbital torque depending on the sign of their orbital and spin Hall conductivities, and spin-orbit coupling, which converts orbitals into spins and vice versa. Pt spacers have been shown to increase the orbital torques in light metal systems [19,21]; however, Pt is also a well-known SHE material. A systematic investigation of the enhancement or suppression of spin and orbital currents in materials with different combinations of orbital and spin conductivities is required.

(vi) The spin diffusion in multilayer structures is usually modeled by semiclassical drift-diffusion equations that account for, e.g., spin backflow at interfaces, the spin-orbit torque dependence on the thickness of the nonmagnet, and the spin Hall and unidirectional magnetoresistance [34–39]. The model has not been extended yet to the OHE, which requires the inclusion of the spin-orbital interconversion mediated by spin-orbit coupling.

(vii) The orbital transmission at the NM/FM interface is more sensitive to the interface quality than spins and, hence, to growth conditions and stacking order [7,18]. It is an open question whether the transmission can be described by a single parameter equivalent to the spin-mixing conductance, which we dub orbital mixing conductance.

(viii) The SHE generates dampinglike and fieldlike spin-orbit torques of comparable strength [1,40]. So far, no theoretical or experimental work has determined with certainty the relative magnitude of the two components of the orbital torque. Assessing their strength may help us to understand the mechanism of accumulation, transfer, and conversion of orbitals.

(ix) The OHE has been attributed to an intrinsic scattering mechanism in elements with orbital texture. The analogy with the SHE [41,42] suggests that also extrinsic processes may contribute to the generation of orbital currents. Measuring the orbital Hall efficiency as a function of the element resistivity may reveal extrinsic orbital effects.

(x) The transmission and absorption of spins and orbitals at the interface with an insulating ferromagnet [43], e.g., yttrium iron garnet (YIG), may be fundamentally different since the latter do not interact with the magnetization. Early experiments reported spin pumping effects in YIG/light metal bilayers, but they were interpreted in terms of the inverse SHE [44].

(xi) The generation and accumulation of orbitals at the NM/FM interface can modulate the longitudinal resistance by the combination of direct and inverse OHE, as recently found in Ref. [22]. Compared with the spin Hall magnetoresistance [35], such orbital Hall magnetoresistance may have a different dependence on the type of ferromagnet, its thickness, and the thickness of the nonmagnet.

(xii) Orbital accumulation might also give rise to a unidirectional magnetoresistance, in analogy to the unidirectional spin Hall magnetoresistance [45]. The underlying mechanism, however, would be intrinsically different since orbitals would not directly alter the magnon population, whereas orbital-dependent scattering might contribute to the conductivity in addition to spin-dependent scattering [46]. On the other hand, the injection into the ferromagnet of electrons with finite orbital momentum may induce an additional source of longitudinal magnetoresistance analogous to the anisotropic magnetoresistance [17].

Orbital effects are thus rich and intertwined with spin transport, allowing for additional means to tune the spin-orbit torque efficiency as well as to understand the transport of angular momentum in thin-film heterostructures. In the following, we address points (i)–(vii) listed above. We provide comprehensive evidence for the occurrence of giant OHEs in 3$d$ and 5$d$ transition metals, reveal the interplay of the OHE and SHE in ferromagnets of variable thickness with and without spacer layers, and establish a phenomenological framework to analyze and efficiently exploit the interplay of spin and orbital currents in metallic heterostructures.

## III. EXPERIMENTS

We studied NM/FM and NM/$X$/FM multilayers where NM = Cr, Mn, or Pt, FM = Co or Ni, and $X$ = Gd or Tb. The samples were grown by magnetron sputtering on a SiN substrate, capped with either Ti(2) or Ru(3.5) (thicknesses in nanometers), and patterned in Hall-bar devices by optical lithography and lift-off. All samples have in-plane magnetization. Current-induced spin-orbit torques were quantified by the harmonic Hall voltage method [40] using angle-scan measurements [51]. We detected the first- and second-harmonic Hall voltage while applying an alternate current with 10 Hz frequency and rotating a constant magnetic field in the easy plane of the magnetization [$xy$ plane; see Fig. 2(a)]. The harmonic signals were measured as a function of current amplitude and field strength [Figs. 2(b) and 2(c)]. The second-harmonic resistance depends on the field angle $\phi$ as $R_{xy}^{2\omega} = \Theta_T \cos\phi + \Phi_T(2\cos^3\phi - \cos\phi)$. Here, $\Theta_T$ is the sum of the dampinglike spin-orbit field $B_{DL}$ and contribution from the thermal gradient along $z$, and $\Phi_T$ depends on the fieldlike spin-orbit field $B_{FL}$ and the Oersted field $B_{Oe}$. Thus the analysis of $R_{xy}^{2\omega}$ measured at different magnetic fields allows for the separation of torques and thermal effects, yielding the magnitude of the spin-orbit fields for a given electric field [51]. These spin-orbit fields exert spin and orbital torques on the magnetization $\mathbf{T}_{DL} = M_s B_{DL} \mathbf{m} \times (\mathbf{p} \times \mathbf{m})$ and $\mathbf{T}_{FL} = M_s B_{FL} \mathbf{m} \times \mathbf{p}$, where $\mathbf{p}$ is the net spin polarization direction, $\mathbf{m}$ is the magnetization vector, and $M_s$ is the saturation magnetization. In the following, we consider uniquely $B_{DL}$ since, apart from Ni/Cr and Co/Pt samples, $B_{FL}$ was too small to distinguish from the Oersted field. The difficult detection of $B_{FL}$ in our samples originates from the very small planar Hall coefficient (of the order of 1 m$\Omega$) to which $\Phi_T$ is proportional. To compare samples with different elements, thickness, and stacking order, we converted $B_{DL}$ into a spin-orbital

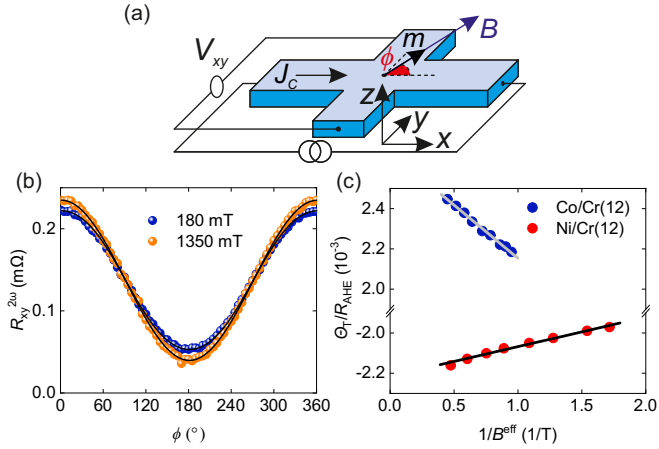

FIG. 2. (a) Schematic of the harmonic Hall effect measurements. An alternating current $J_c$ flows along the Hall bar and generates transverse first- and second-harmonic Hall signals that depend on the angle $\phi$ relative to $x$ of a magnetic field $B$ of constant amplitude. (b) Representative second-harmonic Hall resistance measured in Co(2)/Cr(12) during the rotation of $B$ in the $xy$ plane. The solid lines are fits to the function $R_{xy}^{2\omega} = \Theta_T \cos\phi + \Phi_T(2\cos^3\phi - \cos\phi)$ (see text). (c) Dependence of $\Theta_T$ (dampinglike field $B_{DL}$ + thermal signal) normalized to the anomalous Hall resistance ($R_{AHE}$) on the effective field given by the sum of the applied magnetic field and demagnetizing field. Data are shown for Co(2)/Cr(12) and Ni(4)/Cr(12). The slope of the linear fit (solid lines) is proportional to $B_{DL}$, while its intercept with the vertical axis corresponds to the thermal contribution, which is field independent and can be easily distinguished.

conductivity according to the formula

$$\xi_{LS} = \frac{2e}{\hbar} M_s t_{FM} \frac{B_{DL}}{E}, \qquad (1)$$

where $e$ is the electron charge, $\hbar$ is Planck's constant, $t_{FM}$ is the thickness of the ferromagnet, and $E = \rho J_c$ is the applied electric field ($\rho$ is the longitudinal resistivity, and $J_c$ is the current density) [1,52]. The normalization to the applied electric field avoids the ambiguities intrinsic to the calculation of the current density in a heterostructure. Since in our samples the ferromagnet lies below the nonmagnet, we invert the sign of the measured $B_{DL}$ to follow the convention that Pt has positive spin Hall conductivity. In the literature, Eq. (1) is usually referred to as the *spin Hall conductivity* or *spin-orbit torque efficiency*, which is related to the effective spin Hall angle of the NM layer by $\theta_{LS} = \rho \xi_{LS}$ (Appendix A). Here, we point out that, when the SHE and OHE are considered together, both spin and orbital currents influence $\xi_{LS}$ and their individual quantitative contributions cannot be disentangled because the spin-orbit torques depend on the total nonequilibrium spin angular momentum in the ferromagnet (primary spins + converted spins) but not on the orbital component. This reasoning implies the impossibility of determining separately the spin and orbital Hall conductivities of a material by measuring nonequilibrium effects on an adjacent ferromagnet, even for transparent interfaces. Thus we call $\xi_{LS}$ the spin-orbital conductivity and $\theta_{LS}$ the spin-orbital Hall angle. However, spin and orbital effects can still be distinguished at a qualitative level, as discussed in the following. We also note that a finite OHE could explain, at least in part, the large variability of the spin-orbit torque efficiency found in samples with different ferromagnets, thicknesses, stacking order, and preparation conditions [1].

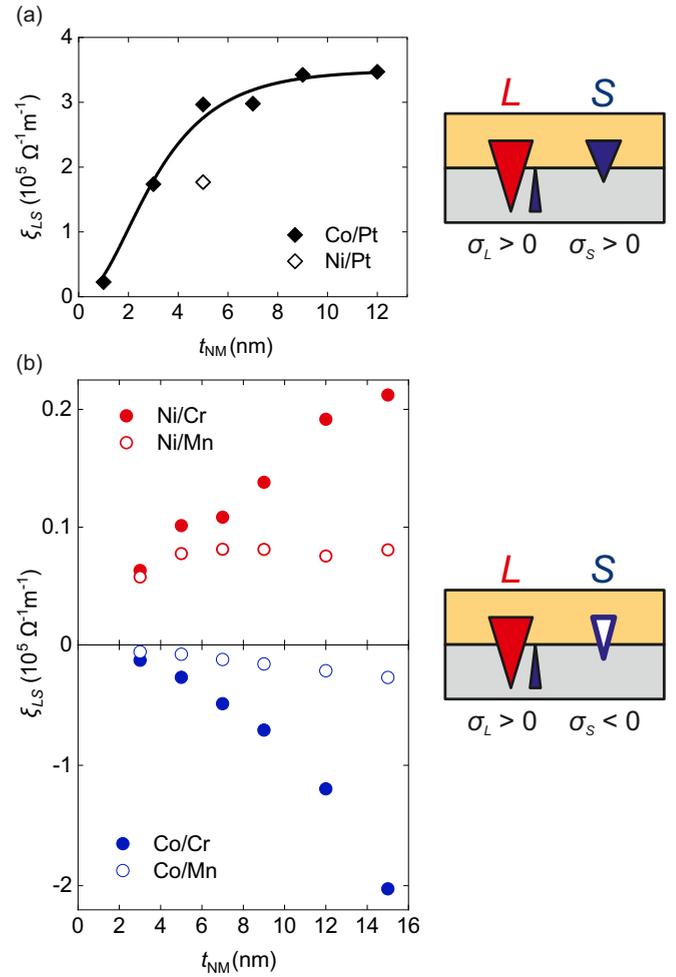

FIG. 3. (a) Spin-orbital conductivity as a function of the thickness of the Pt layer in Co(2)/Pt($t_{NM}$) and in Ni(4)/Pt(5). The solid line is a fit to the drift-diffusion equation [Eq. (2)]. The sign of the spin and orbital Hall conductivities in the nonmagnet is indicated and color-coded in the schematic representing the generation, transmission, and conversion of orbital (red) and positive (blue) or negative (white) spin currents. (b) The same as (a) in FM/Cr($t_{NM}$) and FM/Mn($t_{NM}$), where FM = Co(2) or Ni(4).

## IV. OHE in Cr, Mn, and Pt

### A. Dependence of $\xi_{LS}$ on the thickness of the NM layer

Figure 3 compares $\xi_{LS}$ measured in FM/NM bilayers, where FM is an in-plane magnetized Co(2) or Ni(4) layer and NM is a Cr, Mn, or Pt layer of variable thickness $t_{NM}$. We find that the two 3$d$ light metals generate sizable spin-orbit torques, similar to previous measurements in materials with weak spin-orbit coupling such as V, Cr, and Zr [16,23,44,53]. The torques are remarkably strong in Cr-based samples, for which $\xi_{LS}$ reaches values similar to those for Co/Pt. To the best of our knowledge, this is the highest torque efficiency reported in the literature for a FM/NM bilayer made of light elements. However, the dependence of $\xi_{LS}$ on the type of ferromagnet and on $t_{\rm NM}$ is very different in Cr and Mn with respect to Pt.

TABLE I. Sign of the spin-orbit coupling $\langle \mathbf{L} \cdot \mathbf{S} \rangle$ and orbital and spin Hall conductivity $\sigma_{L,S}$ in selected transition metals (see Refs. [4,11,12,47–49]). A positive orbital (spin) Hall conductivity means that a charge current along $+x$ induces an orbital current (spin current) along $+z$ with orbital (spin) *angular momentum* along $-y$ [50].

|  | Cr | Mn | Co | Ni | Gd | Tb | Pt |
|---|---|---|---|---|---|---|---|
| $\langle \mathbf{L} \cdot \mathbf{S} \rangle$ | − | − | + | + | − | − | + |
| $\sigma_L$ | + | + |  | + |  |  | + |
| $\sigma_S$ | − | − | + | + | − | − | + |

Co/Pt($t_{\rm NM}$) and Ni/Pt(5) have torque efficiencies of comparable magnitude and identical sign. In contrast, when Cr or Mn is used, $\xi_{LS}$ changes sign when Co is replaced with Ni [see Fig. 3(b)]. A comparison between Fig. 3 and Table I indicates that Co/Cr and Co/Mn behave as expected within the framework of the SHE, namely, the sign of the torques is opposite to Co/Pt because the spin Hall conductivity $\sigma_S$ has opposite sign in Cr and Mn relative to Pt. The same argument, however, cannot explain the positive sign of $\xi_{LS}$ in the Ni-based samples since the direction of the spin polarization induced by the SHE is fixed and determined by $\langle \mathbf{L} \cdot \mathbf{S} \rangle_{\rm NM}$. The sign change can be accounted for only by considering the OHE and the opposite sign of the spin and orbital Hall conductivities of Cr and Mn. In this case, the negative $\xi_{LS}$ measured in Co/Cr and Co/Mn indicates that in these samples the spin torque overwhelms the orbital torque. The positive spin-orbital conductivity found with Ni shows instead that the orbital-to-spin conversion in this ferromagnet is so efficient as to make the orbital torque stronger than the spin torque [19]. $\langle \mathbf{L} \cdot \mathbf{S} \rangle_{\rm FM}$ is indeed predicted to be larger in Ni than in Co and positive [7,20,54]; thus a larger amount of the orbital current can be converted into a spin current of opposite sign to the primary spin current generated by Cr or Mn. Therefore the torques exerted on Co are mostly generated outside the ferromagnet thanks to the orbital-to-spin conversion occurring in the nonmagnet. In contrast, the torques on Ni result from the orbital-to-spin conversion inside the ferromagnet.

The variation of $\xi_{LS}$ with the thickness $t_{\rm NM}$ of Cr and Mn is also different from the thickness dependence of the torque efficiency in heavy elements [1,52]. In Co/Pt($t_{\rm NM}$), $\xi_{LS}$ saturates at about 9 nm [Fig. 3(a)]. The fit to the drift-diffusion equation

$$\xi_{LS}(t) = \sigma_{LS}\left[1 - \mathrm{sech}\left(\frac{t}{\lambda}\right)\right] \quad (2)$$

yields a diffusion length $\lambda = 2.2$ nm and an intrinsic spin-orbital Hall conductivity $\sigma_{LS} = 3.5 \times 10^5 \, [\frac{\hbar}{2e}] \, (\Omega\,\mathrm{m})^{-1}$. This value, which agrees with previous works [1,3,41,52,55], assumes a transparent interface and is thus an underestimation of the intrinsic spin-orbital Hall conductivity of Pt. In Cr and Mn, $\xi_{LS}$ increases with $t_{\rm NM}$ and does not saturate, even at $t_{\rm NM} = 15$ nm. The intrinsic spin-orbital Hall conductivity of Cr is thus significantly larger than $\xi_{LS}$ reported in Fig. 3(b). Indeed, fitting $\xi_{LS}$ in Co/Cr($t_{\rm NM}$) with $\lambda$ fixed in the range 15–25 nm yields $5 \times 10^5 < |\sigma_{LS}| < 12 \times 10^5 \, (\Omega\,\mathrm{m})^{-1}$, in good agreement with the predicted giant orbital Hall conductivity of Cr [4].

The trend of $\xi_{LS}(t_{\rm NM})$ hints at two alternatives. The first possibility is that the spin ($\lambda_S$) and orbital ($\lambda_L$) diffusion lengths of Cr and Mn are larger than the typical spin diffusion length of heavy elements. For example, $\lambda_S$ is found to be about 13 and 11 nm in Cr and Mn, respectively, in Ref. [44], whereas $\lambda_S = 1.8$ nm and $\lambda_L = 6.1$ nm in Cr according to Ref. [19]. Alternatively, we argue that it suffices to have a large orbital diffusion length and a nonzero $\langle \mathbf{L} \cdot \mathbf{S} \rangle_{\rm NM}$ for spins to accumulate over long distances, even if the spin diffusion length in the nonmagnet is short (see Sec. VI). Spin torque measurements cannot distinguish between the two possibilities. Nonetheless, the trends in Fig. 3 suggest the possibility to increase the spin-orbital conductivity in FM/Cr samples with large $t_{\rm NM}$ up to and beyond the maximal efficiency of Co/Pt. This possibility has gone unnoticed so far because thin nonmagnetic films ($t_{\rm NM} \approx 5$ nm) are typically considered in torque measurements.

A very long orbital diffusion length in Mn may also explain why $\xi_{LS}$ is smaller in Mn than in Cr at any thickness and independently of the ferromagnet. This result contrasts with theoretical calculations that predict large and comparable orbital conductivities in Cr and Mn [4] but agrees with the spin pumping measurements of Ref. [44]. We notice that Ref. [4] considered the bcc structure to calculate the orbital conductivity of Mn, but different crystalline phases can compete and coexist in Mn thin films [56]. This difference may account for the small experimental value of $\xi_{LS}$. Alternatively, the small spin-orbital conductivity may be determined by a different quality of the FM/Cr and FM/Mn interfaces, to which the orbital current is very sensitive [7,18], possibly because of Co and Mn intermixing [57]. Owing to the larger resistivity of Mn compared with Cr, however, we note that the effective spin-orbital Hall angle of Co(2)/Mn(9) is $\theta_{LS} = -0.03$, which is comparable to $\theta_{LS} = -0.05$ of Co(2)/Cr(9) (Appendix A).

We also notice that interfacial effects (interfacial torques, spin memory loss, and spin transparency) can influence the strength of the torques and hence the spin-orbital conductivity, as shown by the different $\xi_{LS}$ measured in Co/Pt and Ni/Pt samples [58]. However, interfacial effects cannot explain our results, namely, the sign change of $\xi_{LS}$ with the ferromagnet and its monotonic increase with $t_{\rm NM}$, because they should be independent of the thickness of the nonmagnet and become negligible in thick films.

Overall, these measurements provide strong evidence of the OHE and orbital torques in Cr and Mn, in agreement with theoretical predictions and previous studies of Cr-based samples [4,19,25]. Additionally, they show that the spin-orbital diffusion length is much longer in light elements than in Pt, a difference that could be exploited to boost the effective spin-orbital conductivity beyond the limit of FM/Pt samples.

### B. Dependence of $\xi_{LS}$ on the thickness of the FM layer

Theoretical calculations of the spin and orbital transfer at the FM/NM interface predict a different dependence of the spin and orbital torque on the thickness $t_{FM}$ of the ferromagnet [59]. The former is dominant when $t_{FM}$ is small, whereas the orbital torque can be comparable to or larger than the spin torque in thick ferromagnets. As a consequence, the total torque may change sign when $t_{FM}$ increases if $\langle \mathbf{L} \cdot \mathbf{S} \rangle_{NM} \cdot \langle \mathbf{L} \cdot \mathbf{S} \rangle_{FM} < 0$, as, for instance, in the case of Co/Cr. To test this possibility, we measured the torque on the magnetization of Co($t_{FM}$)/Cr(9) and Co($t_{FM}$)/Pt(5) as a function of $t_{FM}$. Figures 4(a) and 4(b) show the torque per unit electric field calculated as $T = M_s \frac{B_{DL}}{E}$. The sign of the torque is opposite in the two sets of samples and does not change in the explored thickness range. This result might suggest that in Co($t_{FM}$)/Cr the orbital torque is always negligible compared with the spin torque. However, a careful analysis indicates a different scenario. After taking into account the dead magnetic layer (0.5 and 0.3 nm in the samples with Cr and Pt, respectively; see Appendix B), we tentatively fit the dependence of $T$ on the ferromagnet thickness to $\sim 1/t_{FM}$. This scaling should reflect the inverse proportionality of the torque amplitude to the magnetic volume when the current-induced angular momentum is generated outside the ferromagnet. In this case, the spin Hall conductivity is constant and solely determined by the charge-to-spin conversion efficiency of the nonmagnet [1], and Eq. (1) yields

$$T = \frac{\hbar}{2e} \frac{\xi_{LS}}{t_{FM}}. \qquad (3)$$

Equation (3) captures well the variation of the torque only for $t_{FM} < 1$–2 nm, in both Co($t_{FM}$)/Cr and Co($t_{FM}$)/Pt. The discrepancy at large thicknesses suggests the presence of a torque mechanism additional to the spin current injection from the nonmagnetic layer. This possibility is corroborated by the thickness dependence of the spin-orbital conductivity, which is different in the two series of samples. In Co($t_{FM}$)/Pt, $|\xi_{LS}|$ is approximately constant up to 3 nm and increases at larger $t_{FM}$ by about 20% [see Fig. 4(c)]. In Co($t_{FM}$)/Cr, instead, $|\xi_{LS}|$ initially increases as the ferromagnet becomes thicker, possibly due to the formation of a continuous Co/Cr interface; then it decreases starting from $t_{FM} = 1$ nm and drops by more than 50% at $t_{FM} = 3$ nm relative to the maximum. Beyond this thickness, it remains approximately unchanged. The distinct thickness dependence in Co($t_{FM}$)/Cr and Co($t_{FM}$)/Pt cannot be ascribed to strain [60] since Co is grown on an amorphous substrate. In addition, strain-induced effects should be similar in the two sample series. Moreover, it cannot be attributed to a variation of the interface quality. Since the latter is expected to improve as Co becomes thicker, the spin-orbital conductivity should increase or remain approximately constant for $t_{FM} > 1$ nm. Furthermore, we exclude that the measured trend depends on uncertainties in the saturation magnetization due to proximity effects since $\xi_{LS}$ depends on the areal magnetization [see Eq. (1)], which is free from ambiguity (see Appendix B). Finally, we rule out self-torques due to the SHE inside the Co layer [49] since control measurements in Co(7)/Ti(3) do not give evidence of torques within the experimental resolution.

Alternatively, we propose that the decrease of the spin-orbital conductivity with $t_{FM}$ in Co/Cr results from the competition between spin and orbital torques in the ferromagnet. As sketched in Figs. 4(d) and 4(e), the spin and orbital currents $J_S$ and $J_L$ decay inside the ferromagnet on a length scale determined by the respective dephasing lengths. In the absence of orbital-to-spin conversion, the spin-orbital conductivity, which depends on the absorption of the injected spin current $\xi_{LS} \sim J_S(0) - J_S(t_{FM})$, increases rapidly with $t_{FM}$ and remains constant afterwards because spin dephasing occurs within a few atomic layers from the interface [61,62]. On the other hand, if we assume that the orbital current is transmitted over a distance longer than its spin counterpart [59] and that part of it is also converted into a spin current $J_{LS}$, then $\xi_{LS} \sim J_S(0) - J_S(t_{FM}) \pm J_{LS}(t_{FM})$ can increase or decrease with $t_{FM}$ depending on the relative sign of $J_S$ and $J_{LS}$, i.e., on the product $\langle \mathbf{L} \cdot \mathbf{S} \rangle_{NM} \cdot \langle \mathbf{L} \cdot \mathbf{S} \rangle_{FM}$. Since the latter is positive (negative) in Co($t_{FM}$)/Pt [Co($t_{FM}$)/Cr], our qualitative model envisages an increase (decrease) of the spin-orbital conductivity with $t_{FM}$, in agreement with our measurements and the thickness dependence predicted in Ref. [59].

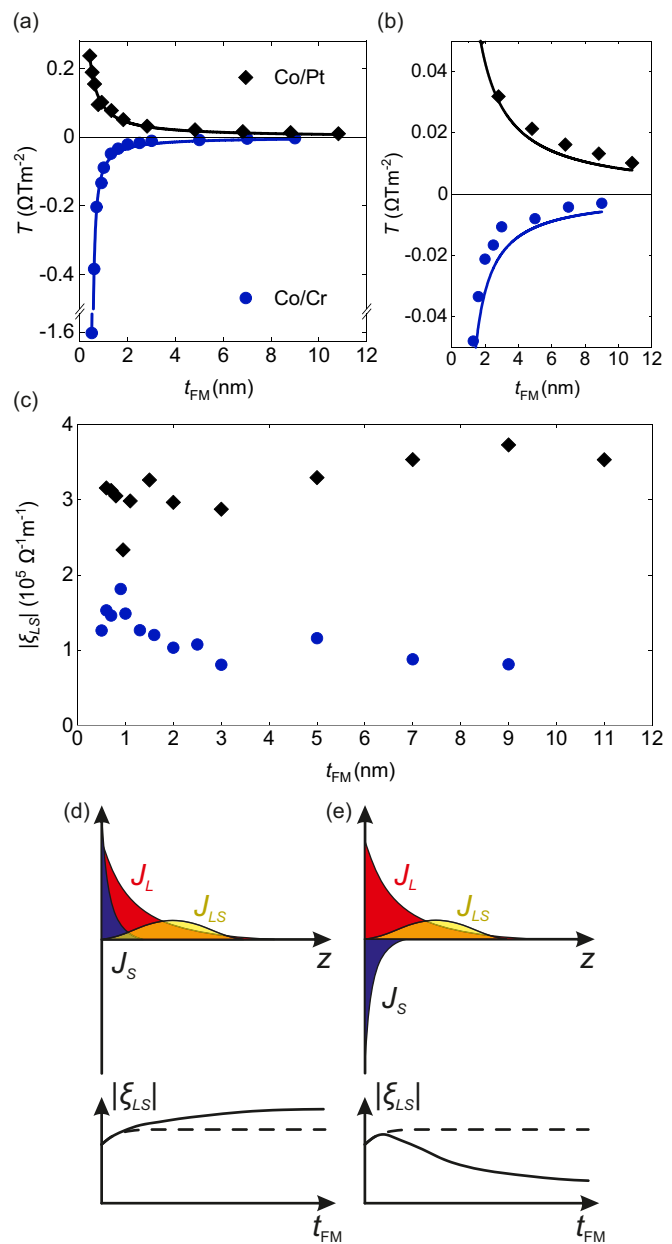

FIG. 4. (a) Dependence of the spin-orbit torque normalized to the applied electric field on $t_{FM}$ in Co($t_{FM}$)/Cr(9) and Co($t_{FM}$)/Pt(5). The solid lines are fits to $\frac{1}{t_{FM}}$. (b) Enlarged view of (a). (c) Dependence of $|\xi_{LS}|$ on $t_{FM}$ in the two sample series. (d) and (e) Schematics showing qualitatively the interplay of the spin $J_S$ and orbital $J_L$ currents, which are injected into the ferromagnet from the interface with the nonmagnetic metal and decay with the distance $z$. Part of the orbital moments is converted into spin moments and generates a spin current $J_{LS}$ with the same (opposite) polarization as the primary spin current in Pt/Co (Cr/Co). $J_S$ yields the spin torque, and $J_{LS}$ yields the orbital torque. The spin-orbital conductivity $\xi_{LS}$ is constant when the orbital-to-spin conversion is negligible (dashed line). It increases with $t_{FM}$ when $J_S$ and $J_{LS}$ add up and decreases when $J_S$ and $J_{LS}$ compete (solid line).

The dependence of *both* $T$ and $\xi_{LS}$ on $t_{FM}$ shows that Co, rather than being a passive layer subject to an externally generated spin current, participates in the overall generation of spin-orbit torques. The active role of the ferromagnet invalidates the assumption on which Eq. (3) rests and explains the deviation of the torque measured at large thicknesses from the $1/t_{FM}$ dependence. Interestingly, these measurements point to a non-negligible OHE in Pt, in accordance with the measurements discussed next.

## V. ORBITAL-TO-SPIN CONVERSION IN A SPACER LAYER

The results presented in Sec. IV show that the spin-orbital conductivity of a light metal can be maximized by a proper choice of the ferromagnet and its thickness. There is, however, a limitation from both a practical and theoretical point of view. According to Hund's third rule, light metals have opposite spin-orbit coupling relative to ferromagnetic Fe, Co, and Ni; thus $\xi_{LS}$ cannot be maximized in such bilayers. As proposed in Sec. II, this optimization may be possible, instead, if the orbital current is converted into the spin current prior to the injection into the ferromagnetic layer [Fig. 1(b)]. This approach requires materials with high spin-orbit coupling between the light metal and the ferromagnet [21]. Although the additional layer can itself be a source of spin current, we show in the following how thickness-dependent measurements reveal the underlying orbital-to-spin conversion and indicate the optimal conversion conditions.

Figure 5 shows the spin-orbital conductivity measured in Co(2)/$X$($t_X$)/Cr(9) and Co(2)/$X$($t_X$)/Pt(5) as a function of the rare-earth thickness, where $X$ is either Gd or Tb. We find a drastic change of the magnitude and sign of the torques upon increasing $t_X$. As the rare-earth layer becomes thicker in Co/$X$($t_X$)/Cr, $|\xi_{LS}|$ first increases, reaching its maximum magnitude at about $t = 3$ nm, and then decreases [notice the negative sign of $\xi_{LS}$ in Fig. 5(a)]. At this thickness, $|\xi_{LS}|$ of Co/$X$(3)/Cr is three to four times larger than in Co/Cr and is thus comparable to or larger than the highest spin-orbital conductivity of Co/Pt [cf. Figs. 3(a) and 5(a)]. In Co/$X$($t_X$)/Pt, instead, $\xi_{LS}$ decreases rapidly with $t_X$, changes sign at 2 nm, and saturates thereafter. This variation, which is similar in samples containing Gd and Tb, is in direct contrast with the widespread assumption that the positive spin Hall conductivity of Pt determines the sign and magnitude of the dampinglike spin-orbit torque in Pt heterostructures.

Indeed, our findings cannot be attributed to the sole SHE in the nonmagnetic layer, nor can they be attributed to the spin-orbit torques generated by the rare-earth layer, which, although present, are too small to explain the sizable change of $\xi_{LS}$ in the trilayers with respect to the Co/Cr and Co/Pt bilayers (see control measurements of Co/Tb and Co/Gd in the Conclusions). Moreover, samples with inverted position of Gd and Tb with respect to the Co layer present spin-orbital Hall conductivities similar to the samples without the spacer, which indicates that the rare-earth layer is not the dominant source of spin-orbit torques (see the Conclusions). Instead, the results in Fig. 5 can be rationalized by considering the combination of OHE, SHE, and orbital-to-spin conversion in the spacer. The net spin current transferred from Cr or Pt to Co depends on the transmission at the interface, the spin and orbital diffusion in the rare-earth layer, and its orbital-to-spin conversion efficiency. Whereas the first two effects always diminish the spin current reaching the ferromagnet, the orbital-to-spin conversion enhances it when $\langle \mathbf{L} \cdot \mathbf{S} \rangle_{NM} \cdot \langle \mathbf{L} \cdot \mathbf{S} \rangle_X > 0$ and weakens it when $\langle \mathbf{L} \cdot \mathbf{S} \rangle_{NM} \cdot \langle \mathbf{L} \cdot \mathbf{S} \rangle_X < 0$. This is the case for samples containing Cr and Pt, respectively (see Table I). The length scale over which the effect takes place is determined by the combination of the spin and orbital diffusion lengths of Gd and Tb. When the spacer is thin relative to these two lengths, the orbital-to-spin conversion supplies the spin current with more spins than those lost by scattering. On the other hand, spin-flip events become dominant at large thicknesses and decrease the transmitted spin current. The spin-orbital conductivity saturates then to a finite value determined by the SHE of the rare-earth layer, as indicated by the similar $\xi_{LS}$ measured in samples with either Cr or Pt and thick spacers ($t_X \geqslant 6$ nm).

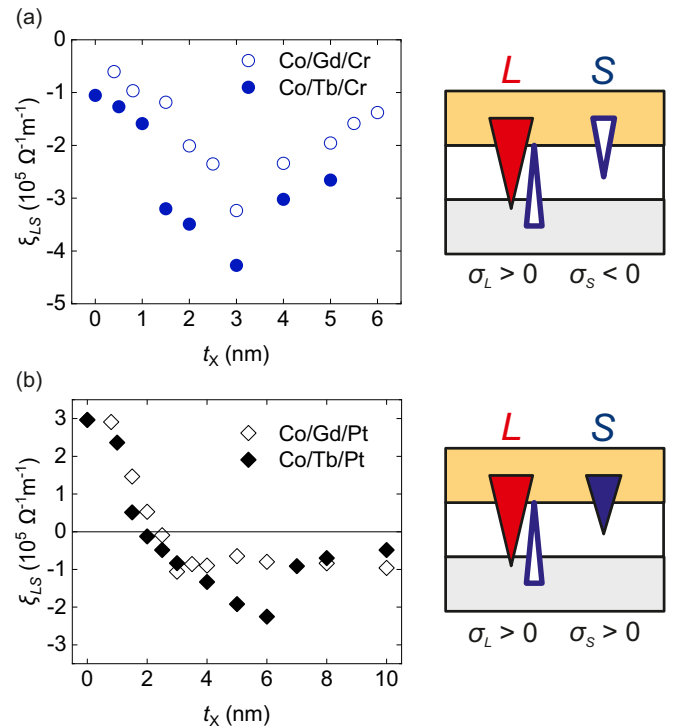

FIG. 5. (a) Dependence of the spin-orbital conductivity on the thickness of the rare-earth spacer in Co(2)/Gd($t_X$)/Cr(9) and Co(2)/Tb($t_X$)/Cr(9). The schematic depicts the conversion of the orbital current into a spin current. Since the spin and orbital Hall conductivities of Cr are opposite and the spin-orbit coupling of Gd and Tb is negative, the primary and converted spin currents have the same sign. (b) The same as (a) in Co(2)/Gd($t_X$)/Pt(5) and Co(2)/Tb($t_X$)/Pt(5). In this case, the primary (blue) and converted (white) spin currents have opposite sign because Pt has positive spin and orbital Hall conductivities and Gd and Tb have negative spin-orbit coupling.

These findings highlight the importance of achieving efficient orbital-to-spin conversion. This can be pursued by sandwiching a rare-earth spacer of optimal thickness between the ferromagnet and the nonmagnet because rare-earth metals are effective enhancers of the conversion but not strong sources of spin-orbit toques [63]. Remarkably, our results also provide evidence of a strong OHE in Pt.

## VI. GENERALIZED DRIFT-DIFFUSION MODEL OF ORBITAL AND SPIN CURRENTS

To shed light on the interplay between spin and orbital currents, we developed a 1D model that takes into account the generation and diffusion of both spin and orbital angular momenta as well as their interconversion mediated by spin-orbit coupling. We consider first a single nonmagnetic layer where an electric field $E$ applied along $x$ induces the SHE and OHE. Let $\mu = \mu_{S,L}$ be the spin or orbital chemical potential and $J_\mu = J_{S,L}$ be the corresponding spin or orbital current along $z$ with spin and orbital polarization along $y$. The generation, drift, and diffusion of spins and orbitals are governed by [34–39]

$$\frac{d^2\mu}{dz^2} = \frac{\mu}{\lambda_\mu^2}, \quad (4)$$

$$J_\mu = -\frac{\sigma}{2e}\frac{d\mu}{dz} + \sigma_H E, \quad (5)$$

where $\lambda_\mu$ is the diffusion length, $\sigma$ is the longitudinal electrical conductivity, and $\sigma_H$ is the spin or orbital conductivity, i.e., the off-diagonal element of the conductivity tensor. Solving these equations yields $\mu = Ae^{z/\lambda_\mu} + Be^{-z/\lambda_\mu}$, with the coefficients $A$ and $B$ obtained by imposing the boundary condition that $J_\mu$ vanishes at the edges of the nonmagnet. In this form, however, the equations of the spin and orbital components are independent and cannot account for the orbital-to-spin and spin-to-orbital conversion mediated by spin-orbit coupling. To capture this process, we add a phenomenological term to Eq. (4) for the spin (orbital) chemical potential that is proportional to its orbital (spin) counterpart, i.e.,

$$\frac{d^2\mu_S}{dz^2} = \frac{\mu_S}{\lambda_S^2} \pm \frac{\mu_L}{\lambda_{LS}^2}, \quad (6)$$

$$\frac{d^2\mu_L}{dz^2} = \frac{\mu_L}{\lambda_L^2} \pm \frac{\mu_S}{\lambda_{LS}^2}, \quad (7)$$

$$J_S = -\frac{\sigma}{2e}\frac{d\mu_S}{dz} + \sigma_S E, \quad (8)$$

$$J_L = -\frac{\sigma}{2e}\frac{d\mu_L}{dz} + \sigma_L E, \quad (9)$$

where the $+$ ($-$) sign corresponds to negative (positive) spin-orbit coupling. Physically, this additional term represents the conversion between spins and orbitals at a rate proportional to the respective chemical potential. Thus, even when the SHE is negligible, a finite spin imbalance is produced in response to the orbital accumulation. The parameter controlling this process is the coupling length $\lambda_{LS}$, which is a measure of both the efficiency and length scale over which the conversion takes place.

We remark that Eqs. (6)–(9) are phenomenological and based on the hypothesis that spin and orbital transport can be described on an equal footing. They assume implicitly the possibility of defining spin and orbital potentials and currents even if the spin and orbital angular momenta are not conserved in the presence of spin-orbit coupling and the crystal field [64]. In this regard, we notice that the spin diffusion model has found widespread use in the quantitative analysis of spin-orbit torques [1,52,65], spin Hall magnetoresistance [35], and surface spin accumulation [55] despite the nonconservation of spin angular momentum. Moreover, there is a fundamental difference between spin and orbital transport that makes the approximations underlying the orbital drift-diffusion model less critical. Contrary to intuition, the crystal field does not quench the *nonequilibrium* orbital moment as efficiently as it suppresses the *equilibrium* orbital moment. This is because the orbital moment is carried by a relatively narrow subset of conduction electron states, namely, its transport is mediated by "hot spots" in **k** space. Since the orbital degeneracy of the hot spots is in general protected against the crystal field splitting, the orbital momentum can be transported over longer distances than its spin counterpart [5,59]. This orbital transport mechanism has no spin equivalent and is supported by the experimental evidence that orbital diffusion lengths in nonmagnets and dephasing lengths in ferromagnets are significantly longer than the corresponding spin lengths, as shown in this paper and in Refs. [19,29]. Further theoretical work is required to ascertain the limits of our spin-orbital model and determine how to capture analytically the spin-orbital interconversion. However, our model is consistent with the Boltzmann approach proposed in Ref. [66] and also reproduces the experimental results, as explained in the following.

To solve the coupled equations (6) and (7), we substitute the former into the latter and obtain

$$\frac{d^4\mu_S}{dz^4} - \left(\frac{1}{\lambda_S^2} + \frac{1}{\lambda_L^2}\right)\frac{d^2\mu_S}{dz^2} + \left(\frac{1}{\lambda_S^2\lambda_L^2} - \frac{1}{\lambda_{LS}^4}\right)\mu_S = 0. \quad (10)$$

The solution to Eq. (10) reads

$$\mu_S(z) = Ae^{z/\lambda_1} + Be^{-z/\lambda_1} + Ce^{z/\lambda_2} + De^{-z/\lambda_2}, \quad (11)$$

where

$$\frac{1}{\lambda_{1,2}^2} = \frac{1}{2}\left[\frac{1}{\lambda_S^2} + \frac{1}{\lambda_L^2} \pm \sqrt{\left(\frac{1}{\lambda_S^2} - \frac{1}{\lambda_L^2}\right)^2 + \frac{4}{\lambda_{LS}^4}}\right] \quad (12)$$

are the combined spin-orbital diffusion lengths that result from the coupling of the spin and orbital degrees of freedom introduced by $\lambda_{LS}$. Equation (11) is the generalization of the standard diffusion of spins valid in the absence of spin-orbital interconversion. Two additional exponentials appear because of the coupling between $L$ and $S$. For the same reason, the spin-orbital diffusion lengths $\lambda_{1,2}$ are a combination of the spin, orbital, and coupling lengths. The same formal solution as Eqs. (11) and (12) holds for the orbital chemical potential because our model treats $\mu_S$ and $\mu_L$ on an equal footing. However, the eight unknown coefficients in Eq. (11) (four for $\mu_S$ and four for $\mu_L$) are in general different between $\mu_S$ and $\mu_L$. They are found by imposing that the spin and orbital currents vanish at the edges of the nonmagnet and that the pair of solutions for $\mu_S$ and $\mu_L$ [Eq. (11)] satisfies Eqs. (6) and (7) at any $z$. Then, we find that the spin chemical potential at the surface of the nonmagnet increases with the thickness $t_{NM}$ as

$$\mu_S(t_{NM}) = 2e\lambda_1 \left( \frac{\sigma_S \mp \frac{\sigma_L}{\lambda_{LS}^2 \gamma_2}}{1 - \frac{\gamma_2}{\gamma_1}} \right) \frac{E}{\sigma} \tanh\left( \frac{t_{NM}}{2\lambda_1} \right)$$
$$+ 2e\lambda_2 \left( \frac{\sigma_S \mp \frac{\sigma_L}{\lambda_{LS}^2 \gamma_1}}{1 - \frac{\gamma_1}{\gamma_2}} \right) \frac{E}{\sigma} \tanh\left( \frac{t_{NM}}{2\lambda_2} \right), \quad (13)$$

where $\gamma_i = \frac{1}{\lambda_i^2} - \frac{1}{\lambda_S^2}$. Equation (13) captures the interplay between the SHE and OHE, which reinforce or weaken each other depending on the sign of the spin-orbit coupling and on $\lambda_{LS}$. In comparison, in the absence of coupling between $S$ and $L$, Eq. (13) would read

$$\mu_S(t_{NM}) = 2e\lambda_S \frac{\sigma_S}{\sigma} E \tanh\left( \frac{t_{NM}}{2\lambda_S} \right), \quad (14)$$

consistent with the standard spin drift-diffusion model. We note that Eqs. (11) and (13) are valid under the condition $\lambda_{LS} > \sqrt{\lambda_S \lambda_L}$ because for smaller values of $\lambda_{LS}$ the solution to Eq. (10) is a linear combination of complex exponential functions, i.e., $\mu_S$ and $\mu_L$ have an oscillatory dependence on $z$. Similar oscillations have been predicted in Ref. [9]. However, we argue that complex solutions to Eq. (12) are incompatible with experimental results since an oscillatory dependence of spin-orbit torques or spin Hall magnetoresistance on the thickness of the nonmagnetic layer has never been observed. The condition $\lambda_{LS} > \sqrt{\lambda_S \lambda_L}$ also implies that the conversion between spins and orbitals cannot occur on a length scale shorter than the shortest distance over which either spins or orbitals diffuse. At the same time, it shows that the conversion is always less efficient than the intrinsic spin and orbital relaxation.

We apply our model to study the interplay of nonequilibrium spins and orbitals induced by the SHE and OHE in two exemplary situations. First, we take a single nonmagnetic layer with negative spin-orbit coupling, e.g., Cr. Figure 6 shows the spin and orbital chemical potentials in three different conditions. In Fig. 6(a), the OHE is turned off ($\sigma_L = 0$), and the SHE is active [$\sigma_S = -10^5$ ($\Omega$ m)$^{-1}$]. In Figs. 6(b) and 6(c), the situation is opposite, namely, $\sigma_L = 10^5$ ($\Omega$ m)$^{-1}$ and $\sigma_S = 0$. In all cases, orbitals (spins) accumulate at the interfaces even if the OHE (SHE) is set to zero. Since $\langle \mathbf{L} \cdot \mathbf{S} \rangle_{NM} < 0$, the two chemical potentials are of opposite sign. When $\lambda_{LS}$ decreases, the spin accumulation resulting from the orbital conversion increases approximately as $\lambda_{LS}^{-2}$ [Fig. 6(b) and Eq. (13)]. Interestingly, we find that even if $\lambda_S$ is small, spins accumulate on a long distance because the spin-orbital diffusion lengths $\lambda_{1,2}$ are dominated by $\lambda_L$ [cf. Figs. 6(b) and 6(c)].

As $\lambda_{LS}$ decreases and the spin-orbit conversion becomes more efficient, both the spin accumulation and the orbital accumulation increase at the sample edges. This effect might seem counterintuitive, because spin-orbit coupling usually induces dissipation of angular momentum. In our model, however, the dissipation of $S$ and $L$ is included in the parameters $\lambda_S$ and $\lambda_L$, respectively, whereas $\lambda_{LS}$ describes the nondissipative exchange of angular momentum between the orbital and spin reservoirs. Thus $\lambda_{LS}$ effectively increases the spatial extent of orbital and spin accumulation. Formally, this happens because one of the two spin-orbital diffusion lengths $\lambda_{1,2}$ increases while the other changes weakly when $\lambda_{LS}$ is reduced. As a consequence, more spins and orbitals can accumulate at the sample edges. This result is similar to the model without spin-orbit coupling, which predicts an increase in $\mu_S$ with the spin diffusion length: $\mu_S(t_{NM} \gg \lambda_S) \sim \lambda_S$ [see Eq. (14)].

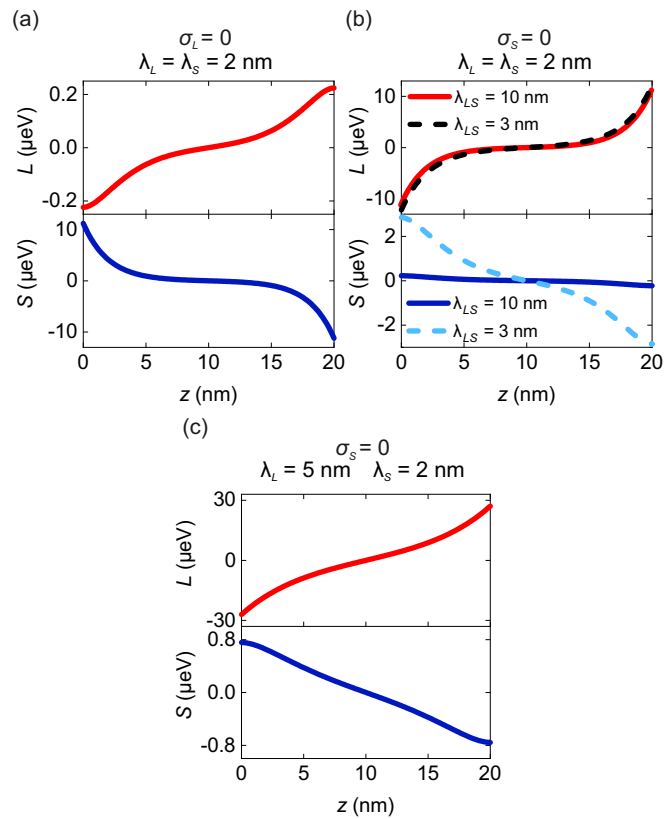

FIG. 6. (a) Orbital and spin chemical potentials in a single nonmagnetic layer with $t_{NM} = 20$ nm, $\sigma_S = -10^5$ ($\Omega$ m)$^{-1}$, $\sigma_L = 0$, $\lambda_S = \lambda_L = 2$ nm, and $\lambda_{LS} = 10$ nm. (b) The same as (a) with $\sigma_S = 0$, $\sigma_L = 10^5$ ($\Omega$ m)$^{-1}$, $\lambda_S = \lambda_L = 2$ nm, and $\lambda_{LS} = 3$ or 10 nm. (c) The same as (a) with $\sigma_S = 0$, $\sigma_L = 10^5$ ($\Omega$ m)$^{-1}$, $\lambda_S = 2$ nm, $\lambda_L = 5$ nm, and $\lambda_{LS} = 10$ nm. In all cases, the resistivity of the NM layer was set to $\rho = 56 \times 10^{-8}$ $\Omega$ m as measured for Cr, and an electric field $E = 5 \times 10^4$ V/m was considered.

Next, we consider a trilayer structure representative of the samples Co(2)/X($t_X$)/Cr(9) and Co(2)/X($t_X$)/Pt(5). We model the spatial variations of $\mu_S$ and $\mu_L$ by four equations of the same type as Eq. (11), two for the rare-earth layer and two for Cr (or Pt). We assume that $\mu_S$, $\mu_L$, $J_S$, and $J_L$ are continuous at the $X$/NM interface ($z = t_X$) and impose the constraint that $J_S$ and $J_L$ relate to $\mu_S$ and $\mu_L$, respectively,

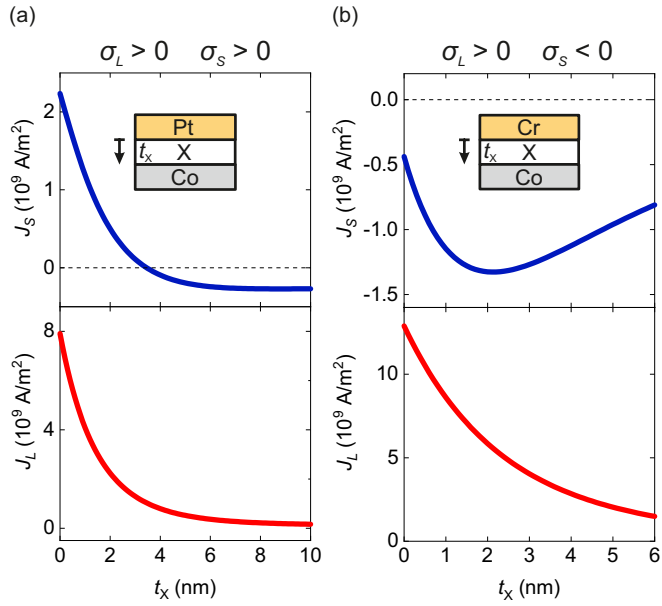

FIG. 7. (a) Calculated spin and orbital currents at the FM/$X$ interface as a function of $t_X$ in Co(2)/$X(t_X)$/Pt(5). (b) The same as (a) for Co(2)/$X(t_X)$/Cr(9). The orbital and spin Hall conductivities of Pt and Cr are indicated above the graphs. The parameters used to calculate the orbital and spin currents can be found in Table II (Appendix C).

through the mixing conductance $G_{S,L}$:

$$J_S^X(t_X) = \frac{G_S^X}{e}\mu_S(t_X), \quad (15)$$

$$J_L^X(t_X) = \frac{G_L^X}{e}\mu_L(t_X). \quad (16)$$

In doing so, we introduce the orbital equivalent of the spin mixing conductance, which is expected to depend on the spin-orbit coupling of the ferromagnet and to influence the strength of the orbital torque. Thus, in our model, $G_L$ takes into account the additional orbital-to-spin conversion occurring in the ferromagnet or at the interface. Furthermore, we only consider the real part of $G_{S,L}$ since the fieldlike torque is small in our samples. Finally, we assume a finite SHE in both the nonmagnetic and rare-earth layers, but smaller in the latter, whereas the OHE is present only in the nonmagnet (see Appendix C for a list of the parameters). We set $\sigma_S > 0$ in Pt, $\sigma_S < 0$ in Cr and in the spacer, and $\sigma_L > 0$ in both Cr and Pt. The spin-orbit coupling is assumed positive in Pt and negative in Cr and in the rare-earth layer. With these reasonable assumptions, we can reproduce qualitatively the results of Fig. 5, namely, the enhancement of the spin-orbital conductivity upon insertion of a rare-earth spacer between Co and Cr and the sign change of the torques when the same layer is sandwiched between Co and Pt. Figure 7 shows the calculated spin and orbital currents, to which spin-orbit torques are proportional, that reach the FM/$X$ interface as a function of the rare-earth thickness $t_X$. In both the case of Cr and the case of Pt the orbital current decreases monotonically as $t_X$ increases because of the orbital diffusion away from the $X$/NM interface. In contrast, the spin current varies differently with $t_X$ depending on whether Cr or Pt is chosen because the primary spin current and the current obtained upon orbital-to-spin conversion in the rare-earth element have the same sign with Cr and have opposite sign with Pt. Thus calculations based on a generalized drift-diffusion model confirm the interpretation of the data in Fig. 5, which cannot be explained without the inclusion of the OHE. We believe that a better quantitative agreement with the measurements could be obtained by including additional effects that we have disregarded, namely, the interfacial resistance ($\mu_S$ and $\mu_L$ not continuous), the interfacial spin and orbital scattering ($J_S$ and $J_L$ not continuous), and the thickness dependence of the spacer resistivity and, possibly, of the diffusion lengths. The model could be extended to account for the orbital conversion in the ferromagnet, which is hidden here behind the orbital mixing conductance. Finally, it may be employed to investigate other transport effects such as the spin Hall magnetoresistance and its orbital counterpart.

## VII. CONCLUSIONS

Our measurements of spin-orbit torques in FM/NM and FM/$X$/NM multilayers with light and heavy metals provide comprehensive evidence for strong OHE effects in 3$d$ and 5$d$ metals and establish a systematic framework to analyze and efficiently exploit the interplay of spin and orbital currents. Owing to the entanglement of the orbital and spin degrees of freedom in materials with finite spin-orbit coupling, this interplay is best described by combined spin-orbital conductivity ($\xi_{LS}$) and diffusion length ($\lambda_{LS}$) parameters rather than by considering the OHE and SHE as two separate effects. The experimental values of $\xi_{LS}$ for the different systems and control samples are summarized in Fig. 8. Corresponding values of the spin-orbital Hall angle $\theta_{LS}$ are reported in Appendix A. We found strong spin-orbit torques produced by the light elements Cr and Mn, whose sign depends on the adjacent ferromagnet, in contrast with torques generated by the SHE. The spin-orbital conductivity increases with the thickness of the light metal layer without indications of saturation. This trend is compatible with spin-orbital diffusion lengths $\lambda_{LS} \gtrsim 20$ nm in these elements and extrapolates to a giant intrinsic spin-orbital conductivity as predicted by theory [4]. Because of the competition between spin and orbital torques, the spin-orbital conductivity varies with the thickness of the ferromagnet in a monotonic or nonmonotonic way depending on the relative sign of $\langle \mathbf{L} \cdot \mathbf{S} \rangle_{NM}$ and $\langle \mathbf{L} \cdot \mathbf{S} \rangle_{FM}$. Furthermore, we show that the interplay between spin and orbital torques can be drastically enhanced by inserting a 4$f$ spacer layer between the nonmagnet and the ferromagnet. As summarized in Fig. 8, the inclusion of a Tb (Gd) spacer results in a fourfold (threefold) increase of the torques generated by Cr and Mn that cannot be attributed to spin currents generated by the rare-earth element. Instead, the enhancement results from the conversion of the orbital current into a secondary spin current of the same sign as the primary spin current. The orbital-to-spin conversion has a striking effect in Pt, when the primary spin current generated by the SHE and the secondary spin currents generated by the OHE interfere destructively. This effect results in the reversal of the spin-orbit torque generated by Pt when the orbital-to-spin conversion rate is stronger than the primary spin current. These findings indicate the presence of a strong OHE and SHE in Cr, Mn, and Pt and highlight the importance of orbital-to-spin conversion phenomena in different types of heterostructures. The largest $\xi_{LS} = -4.3 \times 10^5$ $(\Omega\,\text{m})^{-1}$ and $\theta_{LS} \approx 0.25$ are found in Co(2)/Gd(3)/Cr(9) and Co(2)/Tb(3)/Cr(9) layers. Both of these parameters are larger compared with Co/Pt and previous measurements, indicating that optimization of the thickness of $3d$ metal layers and the insertion of $4f$ spacers lead to giant spin-orbital Hall effects and ensuing spin-orbit torques. The fits of $\xi_{LS}$ as a function of thickness indicate that the spin-orbital conductivity of Cr saturates at values of the order of $10^6$ $(\Omega\,\text{m})^{-1}$, in agreement with theoretical estimates [4]. Finally, we propose an extended drift-diffusion model that treats the orbital and spin moment on an equal footing and includes the orbital-to-spin conversion mediated by spin-orbit coupling. The model explains both the monotonic and nonmonotonic behavior of $\xi_{LS}$ observed in the FM/NM and FM/$X$/NM multilayers as a function of thickness and spin-orbit coupling of the constituent layers. It also shows how the spatial profiles of the orbital and spin accumulation are determined by the combined spin-orbital diffusion lengths and spin and orbital mixing conductances. Overall, our results provide a useful framework to maximize the orbital-to-spin conversion efficiency, interpret experimental results, and address open fundamental questions about orbital transport.

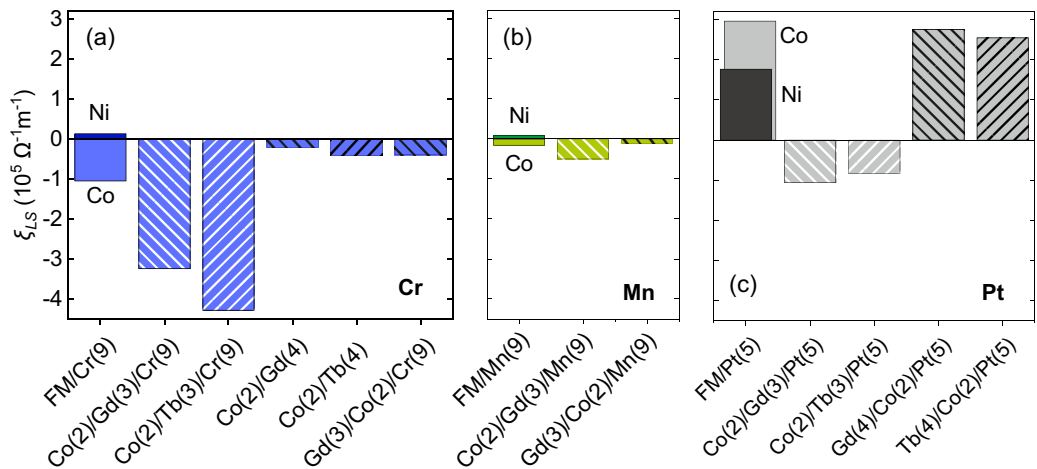

FIG. 8. Comparison of the effective spin-orbital conductivity $\xi_{LS}$ measured in NM/FM and NM/$X$/FM layers where (a) NM = Cr, (b) NM = Mn, and (c) NM = Pt; FM = Co, Ni, $X$ = Gd, Tb. The thickness of each layer is indicated in nanometers in parentheses. The results of control experiments on $X$/FM and NM/FM/$X$ layers are also shown.

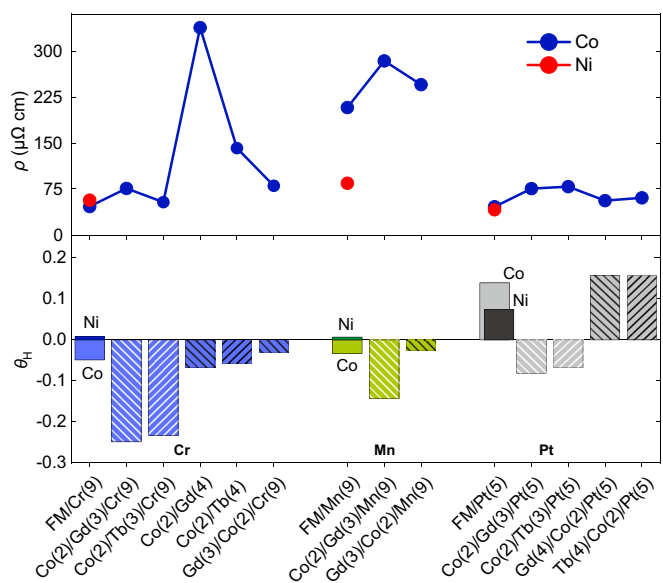

FIG. 9. Resistivity (top) and effective spin-orbital Hall angle (bottom) of the samples in Fig. 8 in the main text.

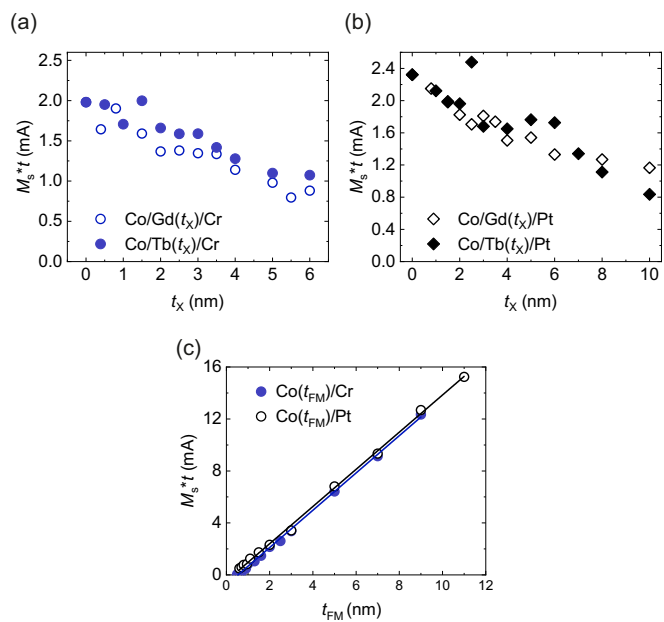

FIG. 10. (a) Dependence of the areal magnetization on the thickness of the rare-earth layer in Co(2)/$X(t_X)$/Cr(9) samples. (b) The same as (a) in Co(2)/$X(t_X)$/Pt(5) samples. (c) Dependence of the areal magnetization on the thickness of the ferromagnet in Co($t_{FM}$)/Cr(9) and Co($t_{FM}$)/Pt(5) samples.

TABLE II. Parameters used in the drift-diffusion model to calculate the spin and orbital currents in a FM/$X$/NM trilayer, where NM is either Cr or Pt. $\lambda_{S,L}$ is the spin or orbital diffusion length, $\lambda_{LS}$ is the spin-orbital conversion length, $\sigma_{S,L}$ is the spin or orbital Hall conductivity, $\alpha = \pm 1$ is the sign of the spin-orbit coupling, $G_{S,L}$ is the spin or orbital mixing conductance, and $\rho$ is the electrical resistivity. The thickness of the Cr (Pt) layer was 9 (5) nm. An electric field $E = 5 \times 10^4$ V/m was considered in both cases.

| | $\lambda_L^{NM}$ (nm) | $\lambda_L^X$ (nm) | $\lambda_S^{NM}$ (nm) | $\lambda_S^X$ (nm) | $\lambda_{LS}^{NM}$ (nm) | $\lambda_{LS}^X$ (nm) | $\sigma_L^{NM}$ [(Ω m)$^{-1}$] | $\sigma_L^X$ [(Ω m)$^{-1}$] | $\sigma_S^{NM}$ [(Ω m)$^{-1}$] | $\sigma_S^X$ [(Ω m)$^{-1}$] | $\alpha_{NM}$ | $\alpha_X$ | $G_L$ [(Ω m$^2$)$^{-1}$] | $G_S$ [(Ω m$^2$)$^{-1}$] | $\rho_{NM}$ (Ω m) | $\rho_X$ (Ω m) |
|---|---|---|---|---|---|---|---|---|---|---|---|---|---|---|---|---|
| Cr | 8 | 2 | 6 | 2 | 20 | 2.5 | $8.2 \times 10^5$ | 0 | $-0.7 \times 10^5$ | $-0.15 \times 10^5$ | $-1$ | $-1$ | $3 \times 10^{14}$ | $10^{14}$ | $56 \times 10^{-8}$ | $115 \times 10^{-8}$ |
| Pt | 1 | 2 | 2 | 2 | 2 | 2.5 | $8.8 \times 10^5$ | 0 | $3.5 \times 10^5$ | $-0.15 \times 10^5$ | $+1$ | $-1$ | $3 \times 10^{14}$ | $10^{14}$ | $33 \times 10^{-8}$ | $115 \times 10^{-8}$ |

## ACKNOWLEDGMENT

We acknowledge the support of the Swiss National Science Foundation (Grant No. 200020_200465).

## APPENDIX A: EFFECTIVE SPIN-ORBITAL HALL ANGLE

Figure 9 shows the effective spin-orbital Hall angle of the samples presented in Fig. 8 in the main text. The Hall angle was calculated according to $\theta_{LS} = \xi_{LS}\rho$, where $\rho$ is the resistivity of the entire stack. However, we refrain from estimating the resistivity of the individual layers and comparing quantitatively $\theta_{LS}$ of the NM layers alone in this way, because the resistivity of the heterostructures depends strongly on interfaces and the thickness of all layers. Similarly to $\xi_{LS}$, we interpret $\theta_{LS}$ as a parameter that describes the simultaneous occurrence of the OHE, the SHE, and orbital-to-spin conversion. The values reported in Figs. 8 and 9 are measured in samples with the thickness specified in the axis labels.

## APPENDIX B: SATURATION MAGNETIZATION

Figure 10 shows the surface saturation magnetization of samples belonging to the series Co(2)/$X(t_X)$/Cr(9), Co(2)/$X(t_X)$/Pt(5), Co($t_{FM}$)/Cr(9), and Co($t_{FM}$)/Pt(5) as a function of the corresponding thickness. The magnetization was measured by superconducting quantum interference device (SQUID) magnetometry on blanket films grown simultaneously to the measured devices. The measurement yields the magnetic moment of the sample, which, after normalization to the sample area, defines the areal saturation magnetization $M_s t_{FM}$. This parameter is to be preferred over the volume saturation magnetization, since the latter depends on the thickness of the ferromagnetically active material. This is in turn difficult to define with certainty in the studied samples because of interdiffusion at interfaces, proximity effects, and possible ferrimagnetic coupling. Such a complexity, however, does not impinge on the calculation of the spin-orbital conductivity because the quantity appearing in Eq. (1) is the areal saturation magnetization $M_s t_{FM}$, not the volume magnetization.

Figures 10(a) and 10(b) show that the areal magnetization decreases upon increasing the thickness of either Gd or Tb in both Cr- and Pt-based samples. We attribute this reduction to the antiferromagnetic interaction between Co and the rare-earth layer. We note that the ferrimagnetic coupling cannot explain our results, namely, the trends presented in Fig. 5. First, the torque efficiency rises by a factor of 3–4 when the thickness of the rare earth $t_X$ increases from 0 to 3 nm, while the areal magnetization decreases only by 20%. Second, the areal magnetization decreases monotonically with $t_X$, while the trends in Fig. 5 are not monotonic with respect to the thickness. For example, the spin-orbital conductivity of Co(2)/$X(t_X)$/Pt(5) saturates in the limit of large $t_X$, whereas the magnetization does not.

The areal magnetization of Co($t_X$)/Cr(9) and Co($t_X$)/Pt(5) samples increases linearly with $t_X$, as expected. The linear fits yield dead layers of about 0.5 and 0.3 nm in the two series, respectively. The dead layer is likely located at the substrate/Co interface and is probably thinner in the Co($t_{FM}$)/Pt(5) series because of proximity effects with Pt. These values have been taken into account in the torque calculation in Fig. 4.

Finally, the saturation magnetization of Co(2)/NM($t_{NM}$) and Ni(4)/NM($t_{NM}$) was found to be independent of $t_{NM}$, except for Co(2)/Pt($t_{NM}$), where the areal magnetization increases by 7% from $t_{Pt} = 1$ nm to $t_{Pt} = 12$ nm (not shown).

## APPENDIX C: PARAMETERS OF THE DRIFT-DIFFUSION MODEL

Table II lists all the parameters used for the calculation of the orbital and spin currents in the Co(2)/$X(t_X)$/Cr(9) and Co(2)/$X(t_X)$/Pt(5) samples (Fig. 7). Some of them have been measured (spin diffusion length of Pt; spin Hall conductivity of Cr and Pt from Co/Cr and Co/Pt samples, respectively; and resistivity). Others have been chosen in accordance with the literature (spin mixing conductance, sign of the spin-orbit coupling, orbital conductivity). The remaining parameters, mostly involving orbitals and the spin-orbital interconversion, are not available in the literature and have been chosen such that the calculations agree qualitatively with the measurements. From this perspective, the extended drift-diffusion model can be used to estimate the order of magnitude of the unknown parameters. For instance, the spin and orbital diffusion length and the spin-orbital coupling length of the rare-earth layer must be of the order of a few nanometers at most for the model to reproduce the measurements in Fig. 5.

[1] A. Manchon, J. Železný, I. M. Miron, T. Jungwirth, J. Sinova, A. Thiaville, K. Garello, and P. Gambardella, Current-induced spin-orbit torques in ferromagnetic and antiferromagnetic systems, Rev. Mod. Phys. **91**, 035004 (2019).

[2] V. E. Demidov, S. Urazhdin, G. de Loubens, O. Klein, V. Cros, A. Anane, and S. O. Demokritov, Magnetization oscillations and waves driven by pure spin currents, Phys. Rep. **673**, 1 (2017).

[3] J. Sinova, S. O. Valenzuela, J. Wunderlich, C. H. Back, and T. Jungwirth, Spin Hall effects, Rev. Mod. Phys. **87**, 1213 (2015).

[4] D. Jo, D. Go, and H. W. Lee, Gigantic intrinsic orbital Hall effects in weakly spin-orbit coupled metals, Phys. Rev. B **98**, 214405 (2018).

[5] D. Go, D. Jo, C. Kim, and H. W. Lee, Intrinsic Spin and Orbital Hall Effects from Orbital Texture, Phys. Rev. Lett. **121**, 086602 (2018).

[6] D. Go and H.-W. Lee, Orbital torque: Torque generation by orbital current injection, Phys. Rev. Research **2**, 013177 (2020).

[7] D. Go, F. Freimuth, J.-P. Hanke, F. Xue, O. Gomonay, K.-J. Lee, S. Blügel, P. M. Haney, H.-W. Lee, and Y. Mokrousov, Theory of current-induced angular momentum transfer dynamics in spin-orbit coupled systems, Phys. Rev. Res. **2**, 033401 (2020).

[8] D. Go, D. Jo, H.-W. Lee, M. Kläui, and Y. Mokrousov, Orbitronics: Orbital currents in solids, Europhys. Lett. **135**, 37001 (2021).

[9] L. Salemi, M. Berritta, and P. M. Oppeneer, Quantitative comparison of electrically induced spin and orbital polarizations in heavy-metal/3$d$-metal bilayers, Phys. Rev. Materials **5**, 074407 (2021).

[10] P. Sahu, S. Bhowal, and S. Satpathy, Effect of the inversion symmetry breaking on the orbital Hall effect: A model study, Phys. Rev. B **103**, 085113 (2021).

[11] T. Tanaka, H. Kontani, M. Naito, T. Naito, D. S. Hirashima, K. Yamada, and J. Inoue, Intrinsic spin Hall effect and orbital Hall effect in 4$d$ and 5$d$ transition metals, Phys. Rev. B **77**, 165117 (2008).

[12] H. Kontani, T. Tanaka, D. S. Hirashima, K. Yamada, and J. Inoue, Giant Orbital Hall Effect in Transition Metals: Origin of Large Spin and Anomalous Hall Effects, Phys. Rev. Lett. **102**, 016601 (2009).

[13] D. Go, D. Jo, T. Gao, K. Ando, S. Blügel, H.-w. Lee, and Y. Mokrousov, Orbital Rashba effect in a surface-oxidized Cu film, Phys. Rev. B **103**, L121113 (2021).

[14] S. Bhowal and S. Satpathy, Intrinsic orbital moment and prediction of a large orbital Hall effect in two-dimensional transition metal dichalcogenides, Phys. Rev. B **101**, 121112(R) (2020).

[15] L. M. Canonico, T. P. Cysne, A. Molina-Sanchez, R. B. Muniz, and T. G. Rappoport, Orbital Hall insulating phase in transition metal dichalcogenide monolayers, Phys. Rev. B **101**, 161409(R) (2020).

[16] Z. C. Zheng, Q. X. Guo, D. Jo, D. Go, L. H. Wang, H. C. Chen, W. Yin, X. M. Wang, G. H. Yu, W. He, H.-W. Lee, J. Teng, and T. Zhu, Magnetization switching driven by current-induced torque from weakly spin-orbit coupled Zr, Phys. Rev. Res. **2**, 013127 (2020).

[17] H. W. Ko, H. J. Park, G. Go, J. H. Oh, K. W. Kim, and K. J. Lee, Role of orbital hybridization in anisotropic magnetoresistance, Phys. Rev. B **101**, 184413 (2020).

[18] J. Kim, D. Go, H. Tsai, D. Jo, K. Kondou, H. W. Lee, and Y. C Otani, Nontrivial torque generation by orbital angular momentum injection in ferromagnetic-metal/Cu/Al$_2$O$_3$ trilayers, Phys. Rev. B **103**, L020407 (2021).

[19] S. Lee, M.-G. Kang, D. Go, D. Kim, J.-H. Kang, T. Lee, G.-H. Lee, J. Kang, N. J. Lee, Y. Mokrousov, S. Kim, K.-J. Kim, K.-J. Lee, and B.-G. Park, Efficient conversion of orbital Hall current to spin current for spin-orbit torque switching, Commun. Phys. **4**, 234 (2021).

[20] D. Lee, D. Go, H.-J. Park, W. Jeong, H.-W. Ko, D. Yun, D. Jo, S. Lee, G. Go, J. H. Oh, K.-J. Kim, B.-G. Park, B.-C. Min, H. C. Koo, H.-W. Lee, O. Lee, and K.-J. Lee, Orbital torque in magnetic bilayers, Nat. Commun. **12**, 6710 (2021).

[21] S. Ding, A. Ross, D. Go, L. Baldrati, Z. Ren, F. Freimuth, S. Becker, F. Kammerbauer, J. Yang, G. Jakob, Y. Mokrousov, and M. Kläui, Harnessing Orbital-to-Spin Conversion of Interfacial Orbital Currents for Efficient Spin-Orbit Torques, Phys. Rev. Lett. **125**, 177201 (2020).

[22] S. Ding, Z. Liang, D. Go, C. Yun, M. Xue, Z. Liu, S. Becker, W. Yang, H. Du, C. Wang, Y. Yang, G. Jakob, M. Kläui, Y. Mokrousov, and J. Yang, Observation of the Orbital Rashba-Edelstein Magnetoresistance, Phys. Rev. Lett. **128**, 067201 (2022).

[23] Q. Guo, Z. Ren, H. Bai, X. Wang, G. Yu, W. He, J. Teng, and T. Zhu, Current-induced magnetization switching in perpendicularly magnetized V/CoFeB/MgO multilayers, Phys. Rev. B **104**, 224429 (2021).

[24] K. V. Shanavas, Z. S. Popović, and S. Satpathy, Theoretical model for Rashba spin-orbit interaction in $d$ electrons, Phys. Rev. B **90**, 165108 (2014).

[25] C.-Y. Hu, Y.-F. Chiu, C.-C. Tsai, C.-C. Huang, K.-H. Chen, C.-W. Peng, C.-M. Lee, M.-Y. Song, Y.-L. Huang, S.-J. Lin, and C.-F. Pai, Toward 100% spin-orbit torque efficiency with high spin-orbital Hall conductivity Pt-Cr alloys, ACS Appl. Electron. Mater. **4**, 1099 (2022).

[26] H. An, Y. Kageyama, Y. Kanno, N. Enishi, and K. Ando, Spin-torque generator engineered by natural oxidation of Cu, Nat. Commun. **7**, 13069 (2016).

[27] T. Wang, W. Wang, Y. Xie, M. A. Warsi, J. Wu, Y. Chen, V. O. Lorenz, X. Fan, and J. Q. Xiao, Large spin Hall angle in vanadium film, Sci. Rep. **7**, 1306 (2017).

[28] J. Bass and W. P. Pratt Jr, Spin-diffusion lengths in metals and alloys, and spin-flipping at metal/metal interfaces: An experimentalist's critical review, J. Phys.: Condens. Matter **19**, 183201 (2007).

[29] Y.-G. Choi, D. Jo, K.-h. Ko, D. Go, and H.-w. Lee, Observation of the orbital Hall effect in a light metal Ti, arXiv:2109.14847.

[30] J.-C. Rojas-Sánchez, N. Reyren, P. Laczkowski, W. Savero, J.-P. Attané, C. Deranlot, M. Jamet, J.-M. George, L. Vila, and H. Jaffrès, Spin Pumping and Inverse Spin Hall Effect in Platinum: The Essential Role of Spin-Memory Loss at Metallic Interfaces, Phys. Rev. Lett. **112**, 106602 (2014).

[31] X. Tao, Q. Liu, B. Miao, R. Yu, Z. Feng, L. Sun, B. You, J. Du, K. Chen, S. Zhang, L. Zhang, Z. Yuan, D. Wu, and H. Ding, Self-consistent determination of spin Hall angle and spin diffusion length in Pt and Pd: The role of the interface spin loss, Sci. Adv. **4**, eaat1670 (2018).

[32] L. Zhu, D. C. Ralph, and R. A. Buhrman, Spin-Orbit Torques in Heavy-Metal–Ferromagnet Bilayers with Varying Strengths of Interfacial Spin-Orbit Coupling, Phys. Rev. Lett. **122**, 077201 (2019).

[33] C. O. Avci, G. S. D. Beach, and P. Gambardella, Effects of transition metal spacers on spin-orbit torques, spin Hall magnetoresistance, and magnetic anisotropy of Pt/Co bilayers, Phys. Rev. B **100**, 235454 (2019).

[34] P. C. van Son, H. van Kempen, and P. Wyder, Boundary Resistance of the Ferromagnetic-Nonferromagnetic Metal Interface, Phys. Rev. Lett. **58**, 2271 (1987).

[35] Y.-T. Chen, S. Takahashi, H. Nakayama, M. Althammer, S. T. B. Goennenwein, E. Saitoh, and G. E. W. Bauer, Theory of spin Hall magnetoresistance, Phys. Rev. B **87**, 144411 (2013).

[36] P. M. Haney, H. W. Lee, K. J. Lee, A. Manchon, and M. D. Stiles, Current induced torques and interfacial spin-orbit coupling: Semiclassical modeling, Phys. Rev. B **87**, 174411 (2013).

[37] V. P. Amin and M. D. Stiles, Spin transport at interfaces with spin-orbit coupling: Formalism, Phys. Rev. B **94**, 104419 (2016).

[38] V. P. Amin and M. D. Stiles, Spin transport at interfaces with spin-orbit coupling: Phenomenology, Phys. Rev. B **94**, 104420 (2016).

[39] K.-W. Kim, Spin transparency for the interface of an ultrathin magnet within the spin dephasing length, Phys. Rev. B **99**, 224415 (2019).

[40] K. Garello, I. M. Miron, C. O. Avci, F. Freimuth, Y. Mokrousov, S. Blügel, S. Auffret, O. Boulle, G. Gaudin, and P. Gambardella, Symmetry and magnitude of spin-orbit torques in ferromagnetic heterostructures, Nat. Nanotechnol. **8**, 587 (2013).

[41] E. Sagasta, Y. Omori, M. Isasa, M. Gradhand, L. E. Hueso, Y. Niimi, Y. Otani, and F. Casanova, Tuning the spin Hall effect of Pt from the moderately dirty to the superclean regime, Phys. Rev. B **94**, 060412(R) (2016).

[42] H. Moriya, A. Musha, S. Haku, and K. Ando, Observation of the crossover between metallic and insulating regimes of the spin Hall effect, Commun. Phys. **5**, 12 (2022).

[43] H. Nakayama, M. Althammer, Y.-T. Chen, K. Uchida, Y. Kajiwara, D. Kikuchi, T. Ohtani, S. Geprägs, M. Opel, S. Takahashi, R. Gross, G. E. W. Bauer, S. T. B. Goennenwein, and E. Saitoh, Spin Hall Magnetoresistance Induced by a Nonequilibrium Proximity Effect, Phys. Rev. Lett. **110**, 206601 (2013).

[44] C. Du, H. Wang, F. Yang, and P. C. Hammel, Systematic variation of spin-orbit coupling with $d$-orbital filling: Large inverse spin Hall effect in 3$d$ transition metals, Phys. Rev. B **90**, 140407(R) (2014).

[45] C. O. Avci, K. Garello, A. Ghosh, M. Gabureac, S. F. Alvarado, and P. Gambardella, Unidirectional spin Hall magnetoresistance in ferromagnet-normal metal bilayers, Nat. Phys. **11**, 570 (2015).

[46] C. O. Avci, J. Mendil, G. S. D. Beach, and P. Gambardella, Origins of the Unidirectional Spin Hall Magnetoresistance in Metallic Bilayers, Phys. Rev. Lett. **121**, 087207 (2018).

[47] K. Ueda, C.-f. Pai, A. J. Tan, M. Mann, and G. S. D. Beach, Effect of rare earth metal on the spin-orbit torque in magnetic heterostructures, Appl. Phys. Lett. **108**, 232405 (2016).

[48] Q. Y. Wong, C. Murapaka, W. C. Law, W. L. Gan, G. J. Lim, and W. S. Lew, Enhanced Spin-Orbit Torques in Rare-Earth Pt/[Co/Ni]$_2$/Co/Tb Systems, Phys. Rev. Applied **11**, 024057 (2019).

[49] W. Wang, T. Wang, V. P. Amin, Y. Wang, A. Radhakrishnan, A. Davidson, S. R. Allen, T. J. Silva, H. Ohldag, D. Balzar, B. L. Zink, P. M. Haney, J. Q. Xiao, D. G. Cahill, V. O. Lorenz, and X. Fan, Anomalous spin-orbit torques in magnetic single-layer films, Nat. Nanotechnol. **14**, 819 (2019).

[50] M. Schreier, G. E. W. Bauer, V. I. Vasyuchka, J. Flipse, K.-i. Uchida, J. Lotze, V. Lauer, A. V. Chumak, A. A. Serga, S. Daimon, T. Kikkawa, E. Saitoh, B. J. van Wees, B. Hillebrands, R. Gross, and S. T. B. Goennenwein, Sign of inverse spin Hall voltages generated by ferromagnetic resonance and temperature gradients in yttrium iron garnet platinum bilayers, J. Phys. D: Appl. Phys. **48**, 025001 (2015).

[51] C. O. Avci, K. Garello, M. Gabureac, A. Ghosh, A. Fuhrer, S. F. Alvarado, and P. Gambardella, Interplay of spin-orbit torque and thermoelectric effects in ferromagnet-normal-metal bilayers, Phys. Rev. B **90**, 224427 (2014).

[52] M. H. Nguyen, D. C. Ralph, and R. A. Buhrman, Spin Torque Study of the Spin Hall Conductivity and Spin Diffusion Length in Platinum Thin Films with Varying Resistivity, Phys. Rev. Lett. **116**, 126601 (2016).

[53] T. C. Chuang, C. F. Pai, and S. Y. Huang, Cr-induced Perpendicular Magnetic Anisotropy and Field-Free Spin-Orbit-Torque Switching, Phys. Rev. Applied **11**, 061005 (2019).

[54] V. P. Amin, J. Li, M. D. Stiles, and P. M. Haney, Intrinsic spin currents in ferromagnets, Phys. Rev. B **99**, 220405(R) (2019).

[55] C. Stamm, C. Murer, M. Berritta, J. Feng, M. Gabureac, P. M. Oppeneer, and P. Gambardella, Magneto-Optical Detection of the Spin Hall Effect in Pt and W Thin Films, Phys. Rev. Lett. **119**, 087203 (2017).

[56] K. Ounadjela, P. Vennegues, Y. Henry, A. Michel, V. Pierron-Bohnes, and J. Arabski, Structural changes in metastable epitaxial Co/Mn superlattices, Phys. Rev. B **49**, 8561 (1994).

[57] G. M. Luo, H. W. Jiang, C. X. Liu, Z. H. Mai, W. Y. Lai, J. Wang, and Y. F. Ding, Chemical intermixing at FeMn/Co interfaces, J. Appl. Phys. (Melville, NY) **91**, 150 (2002).

[58] W. Zhang, W. Han, X. Jiang, S.-H. Yang, and S. S. P. Parkin, Role of transparency of platinum-ferromagnet interfaces in determining the intrinsic magnitude of the spin Hall effect, Nat. Phys. **11**, 496 (2015).

[59] D. Go, D. Jo, K.-W. Kim, S. Lee, M.-G. Kang, B.-G. Park, S. Blügel, H.-W. Lee, and Y. Mokrousov, Long-range orbital transport in ferromagnets, arXiv:2106.07928.

[60] P. Chowdhury, P. D. Kulkarni, M. Krishnan, H. C. Barshilia, A. Sagdeo, S. K. Rai, G. S. Lodha, and D. V. Sridhara Rao, Effect of coherent to incoherent structural transition on magnetic anisotropy in Co/Pt multilayers, J. Appl. Phys. (Melville, NY) **112**, 023912 (2012).

[61] M. D. Stiles and A. Zangwill, Anatomy of spin-transfer torque, Phys. Rev. B **66**, 014407 (2002).

[62] A. Ghosh, S. Auffret, U. Ebels, and W. E. Bailey, Penetration Depth of Transverse Spin Current in Ultrathin Ferromagnets, Phys. Rev. Lett. **109**, 127202 (2012).

[63] N. Reynolds, P. Jadaun, J. T. Heron, C. L. Jermain, J. Gibbons, R. Collette, R. A. Buhrman, D. G. Schlom, and D. C. Ralph, Spin Hall torques generated by rare-earth thin films, Phys. Rev. B **95**, 064412 (2017).

[64] J. Shi, P. Zhang, D. Xiao, and Q. Niu, Proper Definition of Spin Current in Spin-Orbit Coupled Systems, Phys. Rev. Lett. **96**, 076604 (2006).

[65] L. Liu, T. Moriyama, D. C. Ralph, and R. A. Buhrman, Spin-Torque Ferromagnetic Resonance Induced by the Spin Hall Effect, Phys. Rev. Lett. **106**, 036601 (2011).

[66] S. Han, H.-W. Lee, and K.-W. Kim, Orbital Dynamics in Centrosymmetric Systems, Phys. Rev. Lett. **128**, 176601 (2022).